# Stimulated globular scattering and photonic flame effect: new nonlinear optics phenomena

A.A.Esakov, V.S.Gorelik, A.D.Kudryavtseva, M.V.Tareeva, N.V.Tcherniega
P.N.Lebedev Physical Institute, RAS, Leninskii pr., 53, 119991, Moscow, Russia,

## ABSTRACT

Novel nonlinear optical effects - photonic flame effect (PFE)[1,2] and stimulated globular scattering (SGS)[3] - have been discovered. SGS was observed both in forward and backward direction. Pure opal crystal, consisting of the close-packed $SiO_2$ globules with diameter 200 nm, and crystal with pores, filled with molecular liquid, have been studied. Two Stokes components, shifted from the exciting light frequency by 0.4 – 0.6 $cm^{-1}$, have been observed in SGS. Photonic flame effect consisted in the appearance of the few seconds' duration emission in blue-green spectral range under 20 ns ruby laser pulse excitation.

## INTRODUCTION

Photonic crystals have attracted great attention since the first papers concerning such structures[4-6]. Unlike usual crystals such structures have special periodicity of dielectric constant with a period of the electromagnetic wavelength order. One-, two- and three-dimensional photonic crystals exhibit the remarkable properties, which can be effectively used for photon fluxes processing due to the existing of the photonic band gap. Different types of modes defined by the periodical structure of photonic crystal and the possibility of different photonic crystal structure production lead to the important possible applications. The study of the linear optical properties of the photonic band gap have been the task of many theoretical and experimental works and still remain the task to be investigated[7,8]. The description of the electromagnetic field inside the photonic crystal structures (obtained by transfer matrix method[9] or coupled mode theory[10]) gives the clear picture of the transmitted and reflected spectrum, electromagnetic field distribution inside the crystal and their dependence on the parameters of the photonic crystal structure (values of period, number of periods, refractive index contrast). Large values of the electromagnetic field localization in some regions lead to the expectation of the strong enhancement of nonlinear wave-matter interaction in comparison with bulk crystals. Second harmonic generation in different types of photonic crystals was investigated in works[11,12]. Properly chosen photonic crystal exhibits negative refraction at some conditions[13]. Some features of the stimulated Raman scattering in one-dimensional photonic structure was considered[14]. Fully quantum mechanical treatment of the generation of entangled photon in nonlinear photonic crystals at the process of down-conversion was realized in[15]. Photonic band gap properties which are demonstrated by photonic crystals are being actively used for investigation of photon-exciton interaction[16]. Very important kind of 3-D photonic crystals is globular photonic crystal built of globules (balls) with diameter, which may be comparable with visible light wavelength. In nature such crystals exist as mineral - opal, consisting of quartz nanospheres. Space among these spheres (or globules) is filled with different inorganic materials. Recently technology of synthetic opals producing is developed[7,8,17]. Such opals have 3-D periodical structure and are built of ordered close-packed quartz globules with diameter 200-600 nm, organizing 3-D face-centered cubic lattice.

Because the refractive index contrast in opal (ratio $n_{SiO2}/n_{air}$) is about 1,45 the complete photonic band gap in such structure does not exist but the photonic pseudogap takes place. Empty cavities among these globules have octahedral and tetrahedral form. It is possible to investigate both initial opals (opal matrices) and nanocomposites, in which cavities are filled with organic or inorganic materials, for instance, semiconductors, superconductors, ferromagnetic substances, dielectrics, displaying different types of nonlinearities.

Acoustic modes excited in $SiO_2$ balls which compose opal photonic crystal show the effect of phonon modes quantization[18] and are the reason of stimulated globular scattering[3]. Specific features of the acoustic wave propagation in the photonic structures lead to the possibility of the diverging ultrasonic beam focusing into a narrow focal spot with a large focal depth[19].

In the case of the different kinds of molecular liquids infiltrating the opal crystals some types of stimulated scatterings can be observed: stimulated Raman scattering (SRS), stimulated Brillouin scattering (SBS) and others. Peculiarities of spontaneous Raman scattering in photonic crystals has been described in[20]. Spontaneous Brillouin scattering in such structures has been observed in[18]. In this work we present experimental observations of new types of secondary radiation arising in photonic crystals (synthetic opals) under excitation with giant pulses of ruby laser: stimulated scattering in synthetic opals and opal nanocomposites - stimulated globular scattering (SGS) and new nonlinear effect - photonic flame effect (PFE). SGS effect consisted in the appearance of one or two Stokes components, shifted from the exciting light frequency by 0.4 – 0.6 cm$^{-1}$. In PFE we observed few seconds long radiation in blue-green wavelength range under excitation with 20 ns ruby laser pulse. The luminescence could appear with some delay in time in other opal crystals spatially separated with the crystal, illuminated by the laser light.

## 2. EXPERIMENTAL TECHNIQUE AND SOURCES OF EXCITATION

Ruby laser giant pulse (($=694.3$ nm, ($=20$ ns, $E_{max}=0.3$ J) has been used as a source of nonlinear effects excitation in photonic crystals. Exciting light has been focused into the material by lenses with different focus (50, 90, and 150 mm). Sample distance from focusing system and exciting light energy also changed, which gave possibility to make measurements for different power density at the entrance of the sample and for different field distribution inside the sample. Investigations have been fulfilled for different energetical and geometrical conditions. Fabri-Perot interferometers with different bases have been used for SGS spectral structure investigations, which gave possibility to change the range of dispersion from 0.42 to 1.67 cm$^{-1}$. The samples of opal crystals used had the size 3x5x5mm and were cut parallel to the plane (111) .The angle of the incidence of the laser beam on the plane (111) varied from 0 to 60$^0$. Opal crystals consisting of the close-packed amorphous spheres with diameter 200 nm and nanocomposites (opal crystals with voids filled with acetone or ethanol) were investigated. The experimental results did not depend on the angle of incidence of the laser beam. It is necessary to mention that opal matrix, saturated with acetone or ethanol became practically transparent, because refractive indexes of such nanocomposites components (opal matrix and liquid) were near. It allowed observing light scattering in forward direction. PFE has been registered with the help of digital camera. Spectra, observed during PFE investigations, have been studied with the help of mimi-spectrometer with fiber waveguide. Obtained spectra have been compared with the spectra of emission under UV excitation. Light diodes have been used as a source of excitation in this case.

## INVESTIGATIONS OF SGS

### Experimental setup

Principal scheme of the experimental setup for SGS study is shown at the Fig. 1.

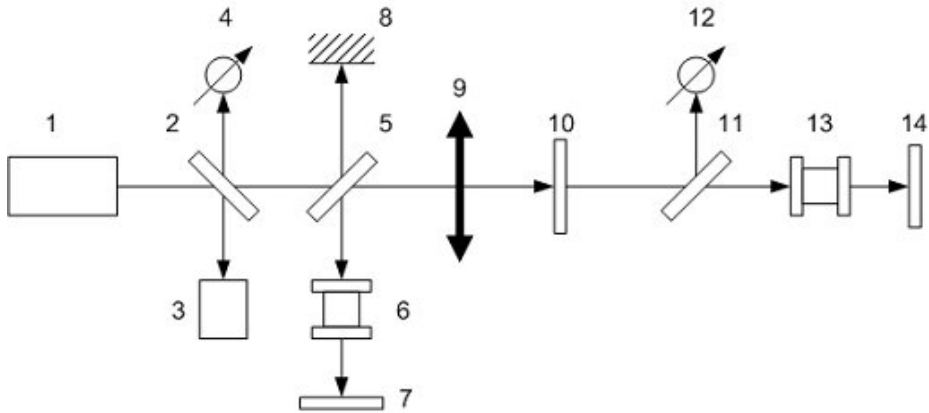

Fig. 1. Setup for SGS study. 1 - ruby laser; 2, 5, 11 – glass plates; 3 – system for laser parameters control; 4, 12 – system for measuring of the scattered light energy in backward and forward direction; 6, 13 – Fabri-Perot interferometers; 7, 14 – system of spectra registration; 8 - mirror; 9 – lens; 10 – photonic crystal.

Ruby laser (1) light has been focused into the photonic crystal (10) with the lens (9). System (3) has been used for pumping light parameters control; systems (4 and 12) measured SGS energy in backward and forward direction. Mirror (8) has been used to register initial light spectrum simultaneously with the scattered light spectrum. In some experiments it was put away. Fig.2 shows interferogram of the ruby laser oscillations. We can observe system of rings, width of which characterizes spectral width of the initial light. For our laser it was 0.015 cm$^{-1}$.

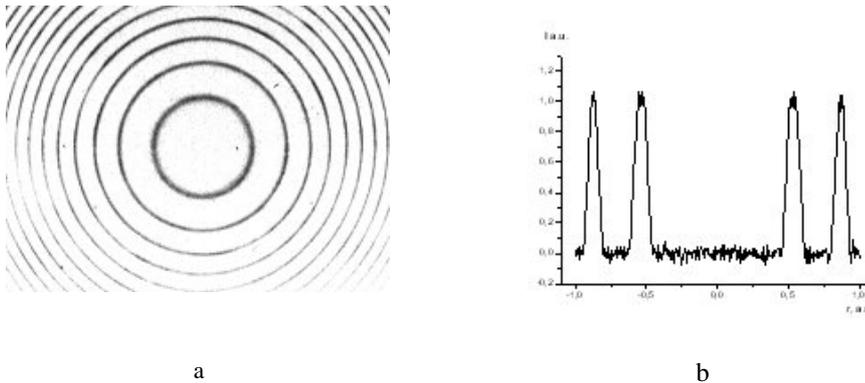

        a                  b

Fig, 2. Interferogram of the exciting laser light; a - picture of the interferogram, b - intensitysity distribution in the spectrum.

### Experimental results

Under ruby laser giant pulse excitation we observed both in opal matrix and in nanocomposites (opal matrix with pores filled with liquids) stimulated scattering of a new type - stimulated globular scattering (SGS). SGS in pure crystal was observed in backward direction (opposite to the initial pumping beam) for pumping light power density more than 0.12 GW/cm$^2$ and its frequency shift was 0.44 cm$^{-1}$. Energy conversion of the laser light into the SGS light was about 0.1 - 0.4. The divergence of the scattered light was of the order 10$^{-3}$ rad. Fig.3a presents interferogam of the scattering from opal matrix. The line width of the scattered light was of the order of the laser light line width.

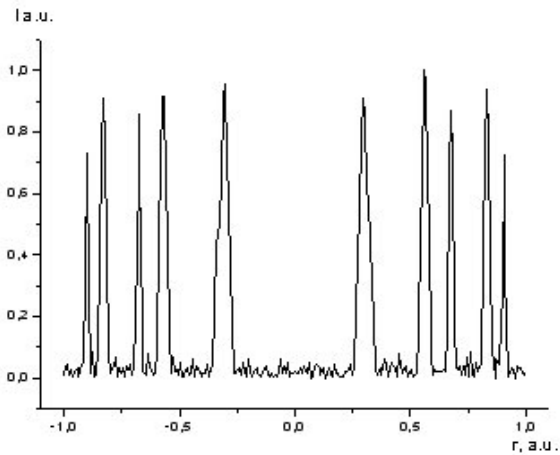

Fig.3a. Interferogram of the SGS in opal in backward direction. Range of Fabri-Perot dispersion is 0.833 cm$^{-1}$.

Two ring systems are observed, which correspond to the pumping laser light (ring of larger diameter) and SGS in backward direction in opal. If we put away mirror 6, reflecting laser light (see Fig.3), only one system of rings is left at the interferogram. This case is shown at the Fig. 3b.

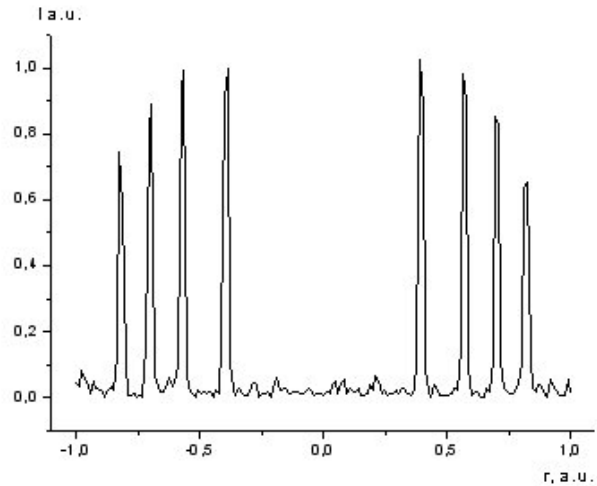

Fig.3b. Interferogram of the SGS in backward direction for the case when mirror 6 is put away

We have also studied SGS in nanocomposites – opal crystals with pores between quartz globules filled with liquids: ethanol or acetone. We registered SGS in such structures both in forward and backward direction. In backward direction for exciting light power more than 0.12 GW/cm$^2$ we observed in opals both with ethanol and acetone first Stokes component with frequency shift 0.4 cm$^{-1}$. Increasing of the pumping power density resulted in the appearance of second Stokes component with frequency shift 0.65 cm$^{-1}$ for acetone and 0.63 cm$^{-1}$ for ethanol. Spectra of backward SGS in acetone are shown at the Figures 4a and 4b.

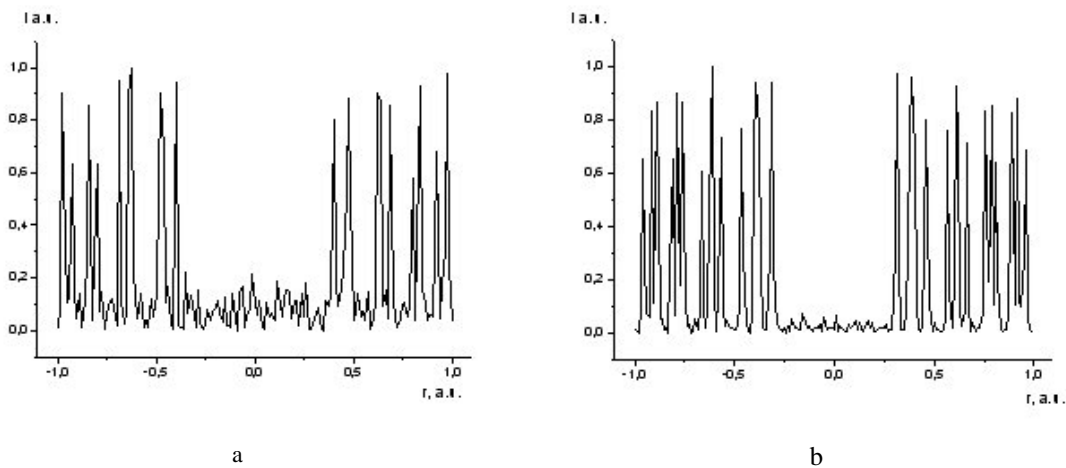

a                                                            b

Fig.4. Backward SGS spectrum for opal filled with acetone. Range of dispersion is 1.67 cm$^{-1}$.
a - pumping power density 0.12 GW/cm$^2$, b - 0.21 GW/cm$^2$.

Larger ring corresponds to the laser light, smaller rings – to the SGS. Backward SGS spectrum in opal, filled with ethanol, is presented at the Fig.5. In this spectrum two Stokes SGS components are observed. Ring, corresponding to the laser light is not observed, because mirror 6 is put away.

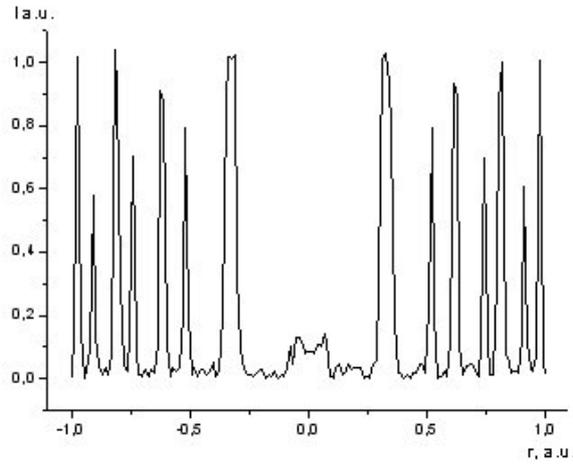

Fig.5. Backward SGS spectrum for opal filled with ethanol for pumping power density 0.21 GW/cm$^2$. Range of dispersion is 1.67 cm$^{-1}$.

In forward direction we observed in opal nanocomposites only one Stokes component of SGS with frequency shift 0.4 cm$^{-1}$ both for ethanol and acetone. Forward SGS spectrum in opal filled with acetone is presented at the Fig.6.

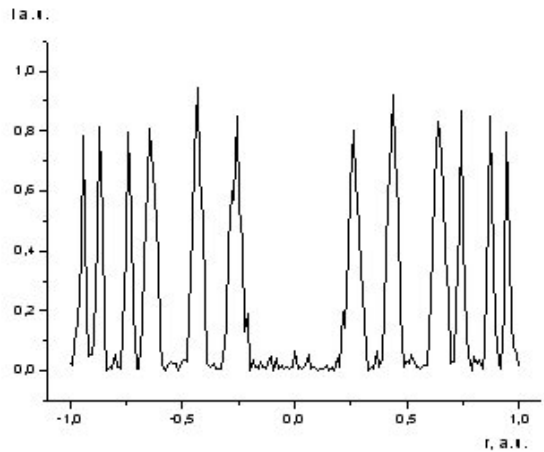

Fig.6. SGS spectrum in forward direction for opal crystal filled with acetone. Range of dispersion is 1.67 cm

Ring with larger diameter corresponds to the laser light, with smaller diameter – to SGS. Spectrum of nanocomposite with ethanol is similar.

### 3.3 Discussion of the results

We compared our experimental results with eigenvalues of quartz globules vibrations. Vibrations of a spherical elastic body have been considered in[21]. They applied for this purpose Lamb's theory dealing with the free vibrations of a homogeneous elastic body of spherical shape under stress-free boundary condition. The equation of motion of such body is

$$\rho \partial^2 \overline{D} / \partial t^2 = (\lambda + \mu)\overline{\nabla}(\overline{\nabla} \cdot \overline{D}) + \mu \overline{\nabla}^2 \overline{D}, \tag{1}$$

where D is the displacement and the parameters ? and ? are Lame's constants. It is possible to solve this equation by introducing scalar and vector potentials. The scalar potential solution of the Helmholtz wave equation is

$$\phi_{si} \propto Z_l(hr) P_l^m(\cos\theta){\cos m\varphi \atop \sin m\varphi} \exp(-i\omega t), \qquad (2)$$

where $Z_l$ is a spherical Bessel function and h = ?/c. The displacement derived from $\phi_{si}$ becomes $\overline{D}_s = \overline{\nabla}\phi_s$. The vector potential is set as $\overline{A} = (r\psi_v, 0, 0)$, where

$$\psi_M \propto Z_l(kr) P_l^m(\cos\theta){\cos m\varphi \atop \sin m\varphi} \exp(-i\omega t) \qquad (3)$$

with k = $\omega/c_l$.

Lamb obtained two types of modes under stress-free boundary condition at spherical surface. The torsional mode is forbidden by selection rules. The eigenvalue equation for spheroidal mode is derived as

$2[\eta^2 + (l-1)(l+2)\{\eta j_{l+1}(\eta)/j_l(\eta) - (l+1)\}] \xi j_{l+1}(\xi)/\xi j_l(\xi) - 0.5\eta^4 + (l-1)(2l+1)\eta^2 + \{\eta^2 - 2l(l-1)(l+2)\} \eta j_{l+1}(\eta)/j_l(\eta) = 0$ (4)

where eigenvalues are

$$\xi = hR = \omega R/c_l = \pi\nu D/c_l \qquad \eta = kR = \omega R/c_t = \pi\nu D/c_t$$

(5)

and $j_l(\eta)$ is the first kind of spherical Bessel function; $c_l$ and $c_t$ are longitudinal and transverse sound velocities.

Equation (9) is solved by setting the parameter $c_l/c_t$. M.H.Kuok, H.S.Lim, S.C.Ng et al [18] calculated frequencies ? for quartz spheres, taking into account values of longitudinal and transverse acoustic mode velocities $c_l$ =5279 m/s and $c_t$ =3344 m/s. Calculated frequencies (in GHz) are following:

$\nu_{10} = 2.617/D, \qquad \nu_{12} = 2.796/D, \qquad \nu_{20} = 4.017/D, \qquad \nu_{30} = 6.343/D,$

(6)

where D is sphere diameter (in $10^{-6}$ m). For our case (D = 200 nm) we have following calculated values of quartz globules frequencies: $\nu_{10}$ = 0.44 cm$^{-1}$, $\nu_{20}$ = 0.68 cm$^{-1}$, which are near to our experimental results. Experimental and calculated values of frequencies are presented in the Table 1.

Table 1. SGS Stokes frequencies for different scattering geometry.

| Scattering geometry | $\nu$, cm$^{-1}$, experimental values | $\nu$, cm$^{-1}$, calculated values |
|---|---|---|
| Backward SGS in opal matrix | 0.44 | 0.44 |
| Backward SGS in nanocomposite (opal with acetone) | 0.40 0.65 | 0.44 0.68 |
| Backward SGS in nanocomposite (opal with ethanol) | 0.39 0.63 | 0.44 0.68 |
| Forward SGS in nanocomposite (opal with acetone) | 0.40 | 0.44 |
| Forward SGS in nanocomposite (opal with ethanol) | 0.37 | 0.44 |

We can see that good agreement exists for pure opal matrix. Liquids, filling pores between quartz globules, may influence their motion. Nevertheless difference between experimental and calculated values even in the case of nanocomposites are not large.

## 4. PHOTONIC FLAME EFFECT INVESTIGATIONS

**Experimental setup**

Experimental arrangement for PFE study is shown at the Fig. 7. Few photonic crystals (synthetic opals) have been placed on the Cu plate, put into the cell with liquid nitrogen. The distance between crystals was from 1 to 5 cm. One of the crystals (3) was illuminated by giant pulse of ruby laser. Luminescence of the crystals was registered by a digital camera and processed at the computer. Spectrum of the luminescence was registered with the help of minipolychromator.

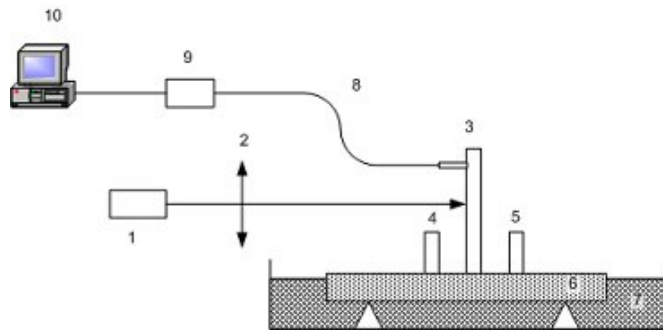

Fig. 7. experimental setup for PPE study. 1- ruby laser; 2- lens; 3, 4, 5 – photonic crystals; 6 – Cu plate; 7 – cell with liquid nitrogen; 8 – fiber wave guide; 9 – minipolychromator; 10 – computer.

### Observation of the photonic flame effect

PFE has been observed at the temperature of liquid nitrogen (77 K). For appropriate cooling opal crystal was placed at the Cu plate, which was partly immersed into liquid $N_2$. Opal crystal has been illuminated with 20 ns pulse of ruby laser. The laser beam was focused to the surface of the crystal . Light spot diameter at the crystal corresponded to the laser beam waist dimensions and was of the order 0.01 mm. Under this excitation few seconds long luminescence in the opal appeared in the blue-green range of spectrum. Dimensions of this shining spot were few mm and varied in dependence of the excitation conditions. The luminescence duration was from 1 to 4 seconds and it looked like inhomogeneous spot changing its spatial distribution and position on the surface of the crystal during this time. PFE has been observed both in opal matrices and in nanocomposites (opals with empty spaces, filled with liquids: acetone or ethanol), Parameters of the secondary emission (duration, threshold) were determined by the geometric characteristic of the illumination and the refractive index contrast of the sample. For optimal geometry of the excitation the power density threshold for opal crystal was 0.12 $Gw/cm^2$, for opal crystals filled with ethanol – 0.05 $Gw/cm^2$, for opal crystal filled with acetone – 0.03 $Gw/cm^2$. Typical secondary emission temporal distribution measured for the part of the crystal displaying the most intensive brightness is shown on the Fig.8. The same behavior is typical for all cases of the secondary emission at these conditions of excitation, but the value of its duration fluctuated from shot to shot (~ 50%). The dependence was close to constant and contained some maximum. Sometimes it was possible to observe few maximums, which were seen by naked eye as light flashes appearing from time to time.

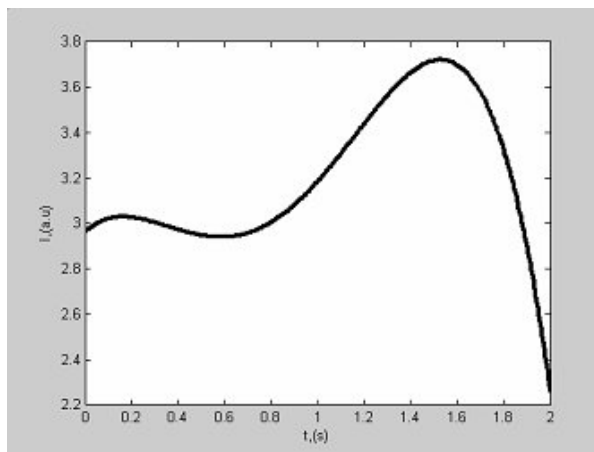

Fig.8. Temporal distribution of the visible luminescence.

Very interesting peculiarity of the effect is possibility of excitation transfer to other crystals, placed at the distance of few cm from illuminated sample. Few opal crystals have been placed on the Cu plate and only one of them was excited by focused laser pulse. When the threshold was reached, this crystal began to emit light. Bright shining of the other crystals began with some time delay after laser shoot. The value of this delay (and the intensity of the luminescence) was determined by the spatial position of the crystals on the plate. The distance between crystals was few cm (up to 5 cm, which was determined by dimensions of the Cu plate). Steel screen placed between crystals did not prevent excitation transfer if distance from lower edge of the screen to the Cu surface was more than 0.5 mm. In order to show the role of the material of the plate used we repeated these experiments with plates of the same size but made of steel and quartz on which opal crystals were placed like in the previous experiments. Luminescence of the same kind in the irradiated crystal took place but the luminescence of the other samples situated on these plates was not observed. Polishing of the Cu plate upper layer (removing of the oxides film) resulted in the threshold increase of energy transfer and decrease of the luminescence intensity. The duration of the other crystals luminescence was of the order of several seconds and temporal behavior was like shown on Fig.8. The typical features of such distribution were existence of maximum and large plateau with near constant value of the intensity.

The effect was also determined by the angle of incidence. For the samples used the value of the angle was chosen experimentally for achieving of the maximal value of the luminescence (it worth to mention that this value differed from 0 and was about $40^0$). Easier the effect was excited in the unprocessed samples. In Fig.9 one can see the secondary emission of the crystals situated at the distance of about 1 centimeter from the crystal which was irradiated.

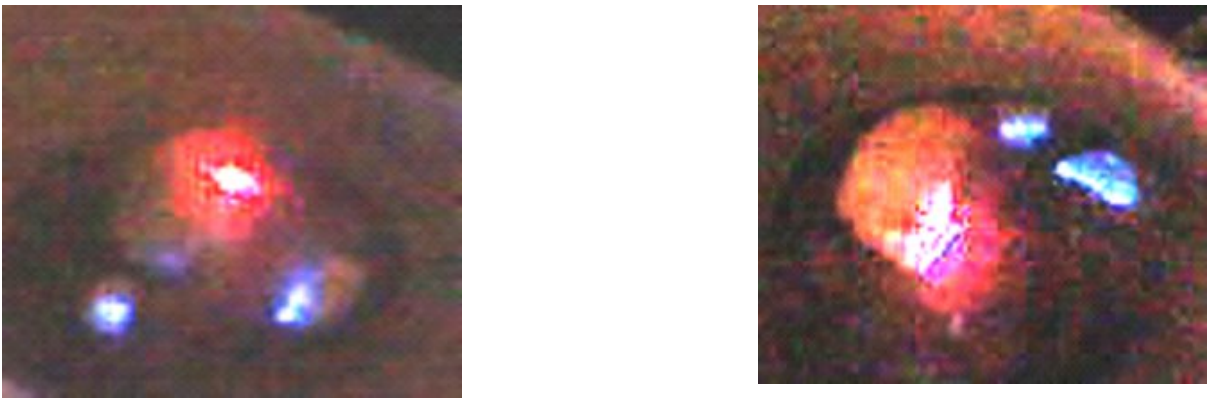

Fig.9. Visible secondary emission of the opal crystals in the case of their irradiating by another (the irradiated crystal can be seen by bright red light; on the left picture it was the crystal in the center, on the right picture it was the crystal on the left). Left picture corresponds to the case where crystals are infiltrated by acetone. Right picture corresponds to the case of the opal crystals without infiltration.

### Spectrum of the secondary emission observed in PFE

Secondary emission spectrum observed in photonic flame effect has been investigated with the help of setup shown at the Fig.7. Spectra of the light emitted by photonic crystal for different pumping light power density are shown at the Fig. 10 (a and b)

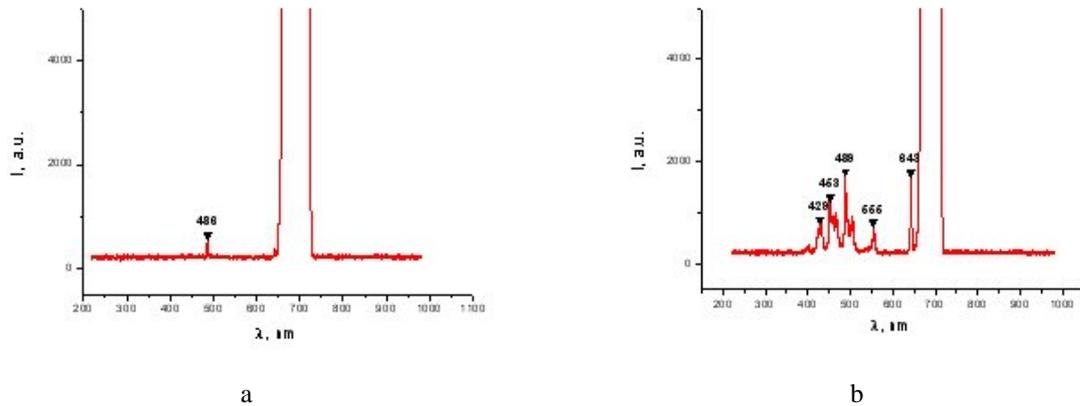

a                          b

Fig. 10. Secondary emission spectrum of a photonic crystal for different laser light power density: a - I = 0.12 GW/cm$^2$, b - I = 0.14 GW/cm$^2$.

Spectrum consisted of the sharp lines with wavelengths: 429.0, 453.0, 489.0, 555.0, 643.0 nm, which corresponds to the antistokes spectral range for exciting line 694.3 nm. Lines intensity in the spectrum strongly depended on the laser pumping intensity, which was evidence of stimulated type of the radiation emission. Maximums observed except line 643.0 nm correlate well with opal spectrum secondary emission under UV excitation, which is evidence of multiphoton character of the observed light emission. In this connection it is interesting to compare spectra of radiation arising in PFE with secondary emission spectra of synthetic opals excited by UV.

### Setup and methodic for study of the opal secondary emission spectra excited by UV radiation

Spectra of secondary radiation in amorphous quartz were studied earlier at excitation by short-wave electromagnetic radiation in view of the big width of the forbidden zone of this material. As it was found out, at excitation by a line 255.3 nm (the second harmonic of a line of generation of the laser on Cu vapour) in a spectrum of a photoluminescence of amorphous quartz peak is present in the region of 300 nm and rather weak maximum close to 380 nm.

In the present work broadband radiation of semi-conductor light diodes with maxima in the region 363.5 and 381.5 nm and average power 30-40 mW has been used as a source of secondary radiation excitation. Usual photoluminescence in the amorphous quartz, caused by presence of defects in it, should not exist in studied samples of artificial opals at the chosen sources of exciting radiation.

The scheme of the used experimental setup for the analysis of luminescence spectra is shown on the Fig.11. Radiation from LED with the help of an optical waveguide of 2 mm diameter was brought to a surface of opals plates of 0.7 and 2.0 mm thick. Secondary radiation gathered from a surface opposite to illuminated, with the help of a quartz optical waveguide with cross section diameter of the photoconductive channel, equal 200 microns. Secondary radiation was brought by this optical waveguide to tiny polychromator FSD8, thus the entrance end face of an optical path served as an entrance crack polychromator. The spectrum of secondary radiation was registered with the help of the CCD-matrix being a part of polychromator.

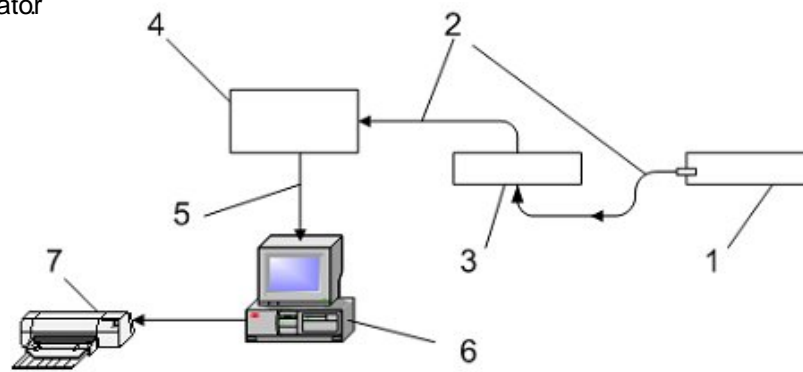

Fig. 11. Setup for the analysis of secondary radiation in artificial opals. 1 - a source of exciting radiation; 2 - a quartz optical path; 3 - a ditch with researched substance; 4 - minipolychromator; 5 - a cable; 6 - a computer; 7 - the printer.

Optical and the spectral characteristics of the polychromator provided high sensitivity of a method. In our experiments time of exposition for the opal secondary emission observation was 0.1-10 s.

### 4.5 Experimental results of opal luminescence spectra under UV excitation

Fig. 12 presents the spectrum of synthetic opal secondary emission. The opal crystal had a form of a plane-parallel plate with thickness of 0,7 mm. Secondary emission was excited by ultra-violet LED with wavelength 363,5 nm. Maxima of a secondary emission thus were observed for wavelengths $\lambda$ =447.3 and 472.4 nm. Intensity of a luminescence was approximately 0.1 from intensity of exciting radiation.

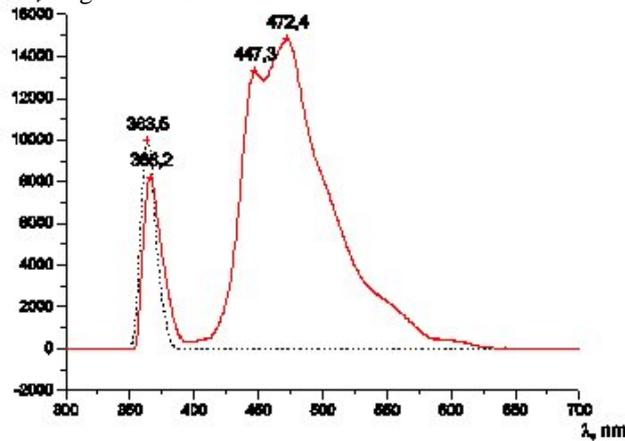

Fig. 12. Dependence of intensity of a luminescence of a plate fell down thickness t = 0,7 mm ($\lambda_0$ =363.5 nm) from length of a wave. The dotted line shows a spectrum of stimulating radiation.

Fig. 13 illustrates secondary emission spectrum of 0.7 mm thick opal plate for excitation by radiation with wavelength 381.5 nm. As one can see, spectra are different for different exciting radiation wavelengths.

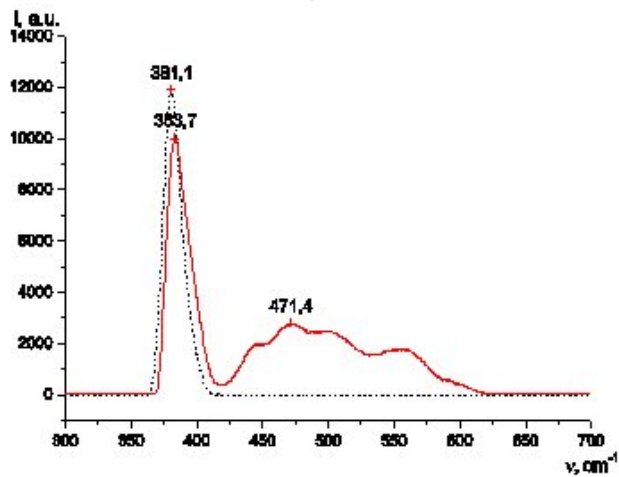

Fig. 13. A spectrum of secondary emission fell down thickness t=0.7 mm at excitation light diode with a maximum of radiation on length of a wave $\lambda_0$ =381.5 nm. The dotted line shows a spectrum of stimulating radiation.

Fig. 14 illustrates similar changes of spectra for opals thickness of 2.0 mm. From spectra comparison it is seen that exciting wavelength change from 363.5 up to 381.5 nm results in increasing of a general width of a secondary emission spectra and number of observed lines.

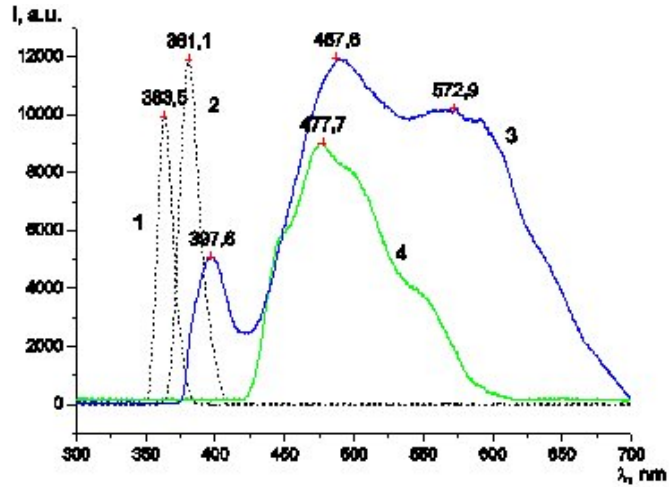

Fig. 14. Comparison of spectra of secondary emission of samples opals (t = 2.0 mm) raised by radiation with lengths of waves $\lambda_0$ =363.5 nm and $\lambda_0$ =381.5 nm. 1, 2 - spectra of stimulating radiation $\lambda_0$ =363.5 and 381.5 nm accordingly; 3-spectrum of secondary emission in disgrace at excitation $\lambda_0$ = 381.5. nm; 4-spectrum of secondary emission in disgrace at excitation $\lambda_0$ =363.5 nm

## 5. CONCLUSIONS

Thus, we discovered experimentally two new nonlinear effects: photonic flame effect and new type of stimulated scattering - stimulated globular scattering. It was observed in globular photonic crystals, created on the base of opal matrices. We observed the effect both in pure synthetic opal crystals and in opal nanocomposites, in which pores between globules are filled with molecular liquids: acetone and ethanol. The effects had threshold character and were was observed for peak power of the ruby laser giant pulse more than $10^7$ W and for lens focus 90 mm.

The main features of PFE are:
    At the excitation of the artificial opal crystal which is placed on the Cu plate at the temperature of the liquid nitrogen by the ruby laser pulse of the nanosecond range long-continued optical luminescence takes place in the case if the threshold of the process is reached;
    In the case of several opal crystals being put on the Cu plate while one of them is being irradiated bright visible luminescence occurs in all samples;
    Temporal behavior and thresholds of the luminescence have been determined. Photonic crystals infiltrated with different nonlinear liquids and without infiltration have been investigated. Investigated transport of the excitation between the samples spatially separated by the length of several centimeters gives the possibility of the practical applications of PFE.

The photonic flame effect can have different explanation. Probably an essential role is played by plasma properties. The slow transport of the excitations from the irradiated crystal to other photonic crystals can be associated with sound waves created due to laser pulse interaction with the sample. Exciton mechanism and surface waveguides on the surface of the Cu plate also can play important role. It was checked that the change of the properties of the plate surface was leading to change of the photonic flame effect. Removing the oxide layer from the plate changed the threshold PFE. Also some analogy with sonoluminescence which inspired the recent studies of nonstationary Casimir effect (see, e.g.[22,23]) deserves a discussion. We will present the details of the photonic flame effect, e.g. spectral characteristics of the light emitted by the photonic crystals in future publication.

Observed frequencies of stimulated globular scattering are in good agreement with eigenvalues of quartz globules, composing opal crystal. Exciting opal matrixes with high-power laser pulses we can expect also appearance of another nonlinear effects, for instance, stimulated Raman scattering, 3-photon and 4-photon parametric processes, nonlinear excited luminescence, optic harmonics excitation, hyper-Rayleigh and hyper-Raman scatterings. We hope that both spontaneous and stimulated globular scattering will give possibility to obtain information about eigenfrequencies spectrum of different materials consisting of spherical nanoparticles including albumen globules, viruses and so on.

## ACKNOWLEDGEMENTS

The authors are grateful to the Russian Foundation for Basic Research for financial support (Grants N 06-02-81024-Bel_ , N 05-02-16205a, N 04-02-16237a).